\documentstyle[preprint,prl,aps]{revtex}

\begin{document}

\draft

\preprint{Submitted to Physical Review Letters}

\author{V. Poduri, D. A. Browne and U. Patil}

\address{
Department of Physics and Astronomy\\
Louisiana State University\\
Baton Rouge, Louisiana 70803}

\date{\today}

\title{Recurrences in Driven Quantum Systems}

\maketitle

\begin{abstract}

We consider an initially bound quantum particle subject to an external
time-dependent field.  When the external field is large, the particle
shows a tendency to repeatedly return to its initial state,
irrespective of whether the frequency of the field is sufficient for
escape from the well.  These recurrences, which are absent in a
classical calculation, arise from the system evolving primarily like a
free particle in the external field.

\end{abstract}


\narrowtext

A very common experimental probe of a physical system is the study of
its response to an imposed time-dependent perturbation.  The behavior
of quantum systems subject to strong time-dependent forces is of
fundamental interest to a variety of situations such as field-enhanced
decay in nuclei\cite{Scully}, highly excited atoms in microwave
fields\cite{Casati} and rf-driven Josephson junctions\cite{Gwinn}.  For
weak external forces, a very successful approach is standard
perturbation theory, where the response of a physical system is
calculated in terms of correlation functions evaluated in the absence
of the external perturbation.  However, there are a steadily increasing
number of experimental situations where the probe is so invasive that
one does not expect perturbative approaches to provide even a
qualitative picture of the behavior of the system.  For example, the
use of high intensity lasers in a variety of experiments during the
last few years has revealed novel phenomena like the decreasing of
ionization probability with increasing laser
intensity\cite{Benvenuto}.  In these situations, there is no simple
theoretical picture akin to perturbation theory and so the analysis of
quantum systems under strong driving remains an unsolved problem.

We consider the following problem: a quantum particle is in the ground
state of a Hamiltonian whose spectrum is partly discrete and partly
continuous.  It is then subjected to a spatially uniform,
time-dependent force.  We are interested in the subsequent behavior of
the system, particularly at strong driving.  We avoid analysis in terms
of quasienergies as is done in the Floquet method\cite{Manakov}, as
that method can only be used if the external forces are purely
periodic.

We solve the time-dependent Schrodinger equation numerically and
monitor the probability for the system to remain in its initial state
as a function of time.  We find that in regions of the parameter space
characterized by strong driving, the system repeatedly exhibits a
tendency to return to its initial state, leading to a series of
recurrences in the initial state occupation probability; this effect is
enhanced if the excitation frequency is above the threshold for
ionization.  These recurrences appear to have no classical analog.

Quantum recurrences in the dynamics of wavepackets in time-{\em
independent} situations have been extensively discussed in the chemical
literature\cite{Marcus}, and have also been studied in the context of
quantum chaos\cite{Heller}.  For time-dependent Hamiltonians, Hogg and
Huberman\cite{Hogg} have established a recurrence theorem:  a system
described by a Hamiltonian with a discrete spectrum subject to a
nonresonant, bounded and time-periodic potential will return
arbitrarily close to its initial configuration infinitely often.  In
our case, the Hamiltonian has a partly discrete and a partly continuous
spectrum and so the above theorem does not apply.

We study a particle of mass $m$ in a one dimensional potential under
the influence of an external force, $F(t)$.  For technical reasons, we
use a vector potential to represent the forcing term.  The
Hamiltonian is
\begin{equation}
{\cal H} = \hbar\omega_0\biggl[
-x_0^2{\partial^2\over \partial x^2} + {V \over \hbar\omega_0} +
{2 \over (\hbar/x_0) } \int F(t) dt\,
\left({x_0\over i}{\partial \over \partial x} \right)
\biggr]
\label{eq:ham1}
\end{equation}
where we have introduced length and frequency scales $x_0$ and
$\omega_0$ with ${\hbar^2/2m} = \hbar\omega_0 x_0^2$, and dropped an
irrelevant time-dependent constant from the right hand side of
Eq.~(\ref{eq:ham1}).  Hereafter, we work in dimensionless units and
choose $F(t) = F_0 \cos(\omega t) $.  The Hamiltonian becomes
\begin{equation}
{\cal H} =
-{\partial^2\over \partial x^2} +  V -
iS {\cal G}(t){\partial \over \partial x}\:,
\label{eq:ham2}
\end{equation}
where $S\equiv 2F_0/\omega$ is a dimensionless measure of the strength
of the forcing term and ${\cal G}(t) = \sin(\omega t) $.

The numerical method that we adopt to solve this problem begins with an
expansion of the wavefunction in a set of time-{\em dependent\/} basis
states $\eta_{k}(t)$:
\begin{equation}
\Psi(t) = \sum_k\, X_{k}(t)\, \eta_{k}(t)
\end{equation}
In order to account exactly for the influence of the external force, we
choose these basis states such that they evolve according to the
field-dependent piece of the Hamiltonian
\begin{equation}
i\frac{\partial \eta_{k}}{\partial t} = -iS {\cal G}(t)
\frac{\partial\eta_{k}}{\partial x}\,.
\end{equation}
The advantage of this procedure is that the choice of the basis at $t=
0$, $\eta_{k}(0)$, is dictated entirely by numerical convenience.  We
choose $\eta_{k}(0) = \exp(ikx)$, so that $\eta_{k}(t) =
\eta_{k}(0)\exp(-iS\gamma(t)k)$ where $\gamma(t) = \int_0^t {\cal
G}(t')dt'$.  These considerations lead to a set of equations for the
amplitudes $X_{k}(t)$:
\begin{equation}
i \dot{X_{k}} = k^2\,X_{k} +
\int\,\frac{dk'}{2\pi}\,{\tilde V}(k-k')\,e^{iS\gamma(t)(k-k')}\,X_{k'}
\label{eq:ampl}
\end{equation}
where ${\tilde V}(k)$ is the Fourier transform of $V(x)$.

For our numerical studies, we choose a smoothly varying potential, $
V(x) = -\lambda\,\text{sech}^2(x)$.  For $\lambda = 0.75$, this
potential admits one bound state
$\psi_{b}(x)=\pi^{-1/2}\,\text{sech}^{1/2}(x)$, with energy
$E_{b}=-0.25$.  We integrate Eq.~(\ref{eq:ampl}) numerically using the
Crank--Nicholson algorithm\cite{recipes} to ensure unitarity and
monitor the probability to return to the ground state,
$P(t)\equiv|\langle\psi(0)|\psi(t)\rangle|^2$.

For low strengths ($S\lesssim0.5$) we find that our numerical results
agree with the predictions of first-order perturbation theory.  For
example, for frequencies above threshold $(\omega>0.25)$, we find that
the probability of remaining in the ground state decays exponentially
with a decay rate given exactly by Fermi's Golden Rule.  Closer
examination of the decay curve shows that superimposed on the
exponential decay is a small periodic ripple whose frequency is twice
that of the external force.  The amplitude of this modulation drops
rapidly as the strength of the external force is decreased.  This small
periodic term arises from coupling between the continuum states induced
by the external force and is dropped in the typical Golden Rule
calculation when the ``rotating wave'' approximation is made.

In the high strength regime the new phenomenon of recurrences appears.
Figure~\ref{mainplot}, which shows the ground state occupation
probability as a function of time for different values of the
excitation frequency, clearly shows that perturbation theory fails to
describe this situation.  While we would intuitively expect rapid and
complete decay, we see a surprising sequence of recurrences in the
initial state occupation probability after nearly complete depletion.
The recurrences become more pronounced at frequencies larger than the
ionization threshold, although we expect that a particle supplied with
sufficient energy to overcome the attractive potential should escape.
Contrary to intuition, the system comes back to its initial state even
though the density of sufficiently energetic quanta is substantial at
these forcing strengths.

We will now focus on the origin of these above threshold recurrences.
At $S = 10$, the potential is clearly overwhelmed by the driving term.
The system behaves largely like a quantum free particle subjected to
the forcing field, the potential being responsible only for minor
corrections to that behavior.  We therefore proceed by first solving
the problem of the free particle in the forcing field exactly and then
adding in the potential through a Born approximation.

The momentum space wavefunction $\tilde{\psi}_f(k,t)$ of a free
particle in the external field can be found from Eq.~(\ref{eq:ham2})
with $V=0$ as
\begin{equation}
\tilde{\psi}_{f}(k,t) = \tilde{\psi}(k,0)\exp[-i(k^2 t+ S\gamma(t)k)],
\label{eq:free}
\end{equation}
where the initial wavefunction, given by the Fourier transform of
$\psi_b(x,0)$, is $\tilde{\psi}(k,0)=(1/\pi\sqrt{2})|\Gamma(1/4 +
ik/2)|^2$.

A first Born approximation to the correct solution (valid in the case
when the potential is small compared to the driving term) is a solution
$\psi_B$ of
\begin{equation}
i{\partial\psi_B\over\partial t} -
[k^2 + S {\cal G}(t) k]\psi_B = V\psi_f\,,
\label{eq:born}
\end{equation}
where the approximation lies in including only $\psi_f$ instead of
$\psi_B$ on the right hand side.  We solve this linear differential
equation for $\psi_B$ numerically and evaluate the approximate ground
state occupation probability.  Figure~\ref{agreeplot}(a) compares the
complete solution with the solution obtained by letting the initial
wavepacket evolve under the influence of the forcing field only, while
Fig.~\ref{agreeplot}(b) shows the improvement gained with the Born
correction.  We see that most of the behavior of the exact solution is
reproduced by the free particle result
$|\langle\psi(0)|\psi_f(t)\rangle|^2$.  The bimodal structures present
in the full problem are missed, but the overall size of the recurrences
is well reproduced.  So the existence of decaying recurrences is
primarily due to the dynamics of wavepacket evolution in the external
field and is relatively independent of the potential.

While the results we have shown here are for a potential with only one
bound state, we have also studied the potential
$V(x)=-6\,\text{sech}^2(x)$, which has two bound states at $E_b=-4$ and
$E_b=-1$. For this problem we see qualitatively similar recurrences at
$S=10$, although with more structure, for excitation frequencies of
$\omega=3$, $4$, and $5.61$.  Note that only the last frequency is
above threshold, and the first one is the resonant frequency for the
transition between the two bound states.

To study the dependence of the recurrences on the form of the initial
state, we examine the behavior of a free particle under the influence
of the external force for an initial state given by a Gaussian
wavepacket, $\psi_G(x)\propto\exp(-x^2/4a^2 + ik_0x)$, of width $a$ and
momentum $k_0$.  The recurrence probability $P_G(t)$ is given by
\begin{equation}
P_G(t) = {2 \over \sqrt {4+t^2/a^4}}
\exp\left[-{(S \gamma(t) + 2k_0 t/a)^2 \over 4+ t^2/a^4}\right] .
\label{eq:gauss}
\end{equation}

In Fig.~\ref{gaussplot} we show the behavior of Eq.~(\ref{eq:gauss})
for $k_0=1$ and $a=1$.  These results are quite similar to those of
Fig.~\ref{agreeplot} and clearly show that recurrences are generic
features of strongly driven quantum systems.  We also see from
Eq.~(\ref{eq:gauss}) that recurrences will appear when the impulse
given the particle in one cycle of the external field $S\gamma(t)$
dominates the characteristic momentum of the particle.  Furthermore,
the decay of the recurrence height is controlled by the prefactor in
Eq.~(\ref{eq:gauss}) which arises from spreading of the wavepacket.  In
general, as we can see from Eq.~(\ref{eq:free}), if the argument of the
exponential is stationary for $k$ values where $\tilde{\psi}_f(k,0)$ is
large, then the wave function has a large overlap with the initial
state and the particle shows recurrences.

One further question that naturally arises is whether these recurrences
have a classical analog.  This is particularly interesting since Davis
and Heller\cite{Davis} have studied wavepacket evolution in several
model potentials in time-independent situations and conclude that there
is indeed very good correspondence between the time average of the
quantum overlap $|\langle\psi(0)|\psi(t)\rangle|^2$ and its classical
analog.  Furthermore, a classical mechanism involving the return of an
electron to the origin which then scatters out a second electron has
been invoked\cite{Corkum} in a model for strong-field double ionization
of atoms\cite{Walker}.

To compare the classical and quantum dynamics, we
consider\cite{Davis,Barb} a cloud of classical particles whose initial
conditions are chosen so that their probability distribution mimics
$|\psi_0|^2$ as much as possible.  Since we have chosen an energy
eigenfunction as an initial state, we have several choices for the
initial distribution of the classical cloud.  We studied two different
classical distributions.  In one, we fixed the position probability
density to match $|\psi(x)|^2$ and determined the momenta from the
fixed energy, giving positive and negative momenta equal weight.  In
the second case, we generated initial conditions with position and
momentum probability densities corresponding to Gaussians; the width of
the initial position distribution was matched with that of the initial
quantum state.  In either case, we could represent the analog of the
quantum overlap by using either the overlap of the time evolved
position probability density with the initial one, $\int
dx\,\rho(x,0)\rho(x,t)$, or an overlap integral\cite{Davis} in phase
space, $\text{Tr}\,[\rho(0)\,\rho(t)]$.

We find that in all cases there are no similarities between the
classical and quantum results; the classical distributions fail to show
recurrences for the same values of the parameters of the forcing term.
We thus conclude that the recurrences discussed here are purely quantum
mechanical in nature, and that the classical model for the recurrence
used in studying the double-ionization of atoms\cite{Corkum}
dramatically underestimates the amplitude for an electron to return to
its original bound state, which may explain the discrepancy between the
classical model and the experimental results\cite{Walker}.

To examine whether the phenomenon of recurrences in strong driving
fields exists in higher dimensional phase spaces, we have also
considered an effective 2D problem: the hydrogen atom in its ground
state subject to linearly polarized radiation.  We observe that
recurrences in the ground state occupation probability do exist at
strong driving above the ionization threshold, even though they are
much smaller.  This work shall be reported elsewhere.

We also expect recurrences even when the external force is not purely
periodic.  The crucial parameter that determines whether the external
force is strong is the coefficient of the momentum operator in the last
term in Eq.~(\ref{eq:ham1}), $\int F(t) dt/(\hbar/x_o)$, which can be
interpreted as the impulse delivered to the system in one cycle in
units of the characteristic momentum of a particle in the potential.
For above threshold excitation, when this number exceeds the strength
of the potential, the system will behave like a free quantum wavepacket
and show all resulting effects described here.

We have therefore shown that whenever a quantum system is subject to a
uniform force varying sinusoidally in time with a strength parameter
that overwhelms the potential, above threshold excitation will lead to
a set of decaying recurrences in the initial state occupation
probability.  While the exact nature of the potential and the shape of
the initial packet determine the detailed behavior and the long time
decay in the magnitude of the recurrences, the existence of the
recurrences is relatively insensitive to these features of the
problem.  While qualitatively similar decaying recurrences have been
observed\cite{Marcus,Heller} in time-independent situations, the ones
we describe are more startling; the quantum system repopulates the
ground state even when more than enough energy is available from the
external field for escape from the well.

This research was supported by the National Science Foundation under
Grant No.~DMR--9020310.

\begin{figure}

\caption{
Probability of initial state occupation as a function of reduced time.
$T=2\pi/\omega$ is the period of the forcing function.  Note that
$\omega/\omega_0 = 0.25$ corresponds to threshold excitation.  }

\label{mainplot}

\end{figure}

\begin{figure}

\caption{
Agreement between approximate and exact calculations for above
threshold excitation.  $T=2\pi/\omega$ is the period of the forcing
function.  The exact calculation is shown in both figures by the solid
line, while the dashed line shows (a) the free particle result
$|\langle\psi(0)|\psi_f(t)\rangle|^2$ from Eq.~(\protect\ref{eq:free})
and (b) the Born approximation $|\langle\psi(0)|\psi_B(t)\rangle|^2$
from Eq.~(\protect\ref{eq:born}).}

\label{agreeplot}

\end{figure}

\begin{figure}

\caption{
Recurrences in the time evolution of a Gaussian wavepacket subject to
an external field, according to Eq.~(\protect\ref{eq:gauss}) with
$k_0=a=1$ for two different driving freqencies.}

\label{gaussplot}

\end{figure}

\end{document}